\documentclass[pra,aps,twocolumn]{revtex4}
\usepackage{graphicx}% Include figure files
\usepackage{dcolumn}% Align table columns on decimal point
\usepackage{amsmath}
\newcommand{\ket}[1]{|#1\rangle}
\newcommand{\bra}[1]{\langle #1|}
\newcommand{\proj}[1]{\ket{#1}\bra{#1}}
\usepackage{bm}% bold math

\begin{document}

% \preprint{APS/123-QED}

\title{Deterministic steady-state subradiance within a single-excitation basis}

\author{Meng-Jia Chu$^{1,3}$}
%\email{renjun@hebtu.edu.cn}
\author{Jun Ren$^{1,2,3}$}
\email{renjun@hku.hk}
%\author{Yan-Kui Bai$^{1,2}$}
%\email{ykbai@semi.ac.cn}
\author{Z. D. Wang$^{1,3}$}
\email{zwang@hku.hk}

\affiliation{$^1$ Department of Physics and HK Institute of Quantum Science $\&$ Technology, The University of Hong Kong, Pokfulam Road, Hong Kong, China\\
$^2$ College of Physics and Hebei Key Laboratory of Photophysics Research and Application, Hebei Normal University, Shijiazhuang, Hebei 050024, China\\
$^3$ Hong Kong Branch for Quantum Science Center of Guangdong-Hong Kong-Macau Great Bay Area, Shenzhen, China}

\collaboration{CLEO Collaboration}%\noaffiliation

\date{\today}% It is always \today, today,

\begin{abstract}
Subradiance shows promising applications in quantum information, yet its realization remains more challenging than superradiance due to the need to suppress various decay channels. This study introduces a state space within a single-excitation basis with perfect subradiance and genuine multipartite quantum entanglement resources for the all-to-all case. Utilizing the quantum jump operator method, we also provide an analytical derivation of the system's steady final state for any single-excitation initial state. Additionally, we determine the approximate final state in the quasi-all-to-all coupling scenario. As an illustrative example, we evaluate the coupling and dynamical properties of emitters in a photonic crystal slab possessing an ultra-high quality bound state in the continuum, thereby validating the efficacy of our theoretical approach. This theoretical framework facilitates the analytical prediction of dynamics for long-lived multipartite entanglement while elucidating a pathway toward realizing autonomous subradiance in atomic systems.

\end{abstract}

\maketitle

%\tableofcontents

\section{Introduction}
Multipartite quantum entanglement is an important quantum resource for quantum information processing in many-body systems \cite{amico08rmp,horo09rmp}. In a three-qubit pure-state system, there are two inequivalent types of genuine tripartite entanglement: the GHZ class and the W class \cite{cirac00pra}. The three-qubit GHZ state $(\ket{000}+\ket{111})/\sqrt{2}$ is considered the maximally entangled state of three qubits, as all the single-qubit reduced states of the GHZ state are entirely mixed. However, under particle losses, the entanglement properties of the state GHZ are extremely fragile. In contrast, the entanglement of W state, $(\ket{001}+\ket{010}+\ket{100})\sqrt{3}$, is maximally robust under loss of any one of the three qubits. In addition, multipartite W-class states play important roles in quantum memory and quantum communication \cite{cirac00pra,zhao07prl}, quantum measurement \cite{gu12nc}, and quantum error correction \cite{briegel01prl,bell14nc,livingston22nc,zhang23prx}. In the framework of single-excitation, the collective states are usually entangled as a W-class, in which a single excitation is dispersed among multiple atoms.

In a collection of atoms, the collective emission rate can exceed $\gamma_0$ (the vacuum emission rate of single excited atom \cite{dirac27prsa}), resulting in a superradiant state \cite{dicke54prl,eberly71pra,haroche82pr}, or be less than $\gamma_0$, leading to a subradiant state \cite{dicke54prl,lehmberg70pra,agarwal74pra,milonni74pra}. While superradiance and subradiance have attracted significant research interest and enabled various applications in recent years \cite{reimann15prl,mh17prl,pog19prl,masson20prl,rob21pra,ruks22prr,rubies22pra,masson22nc,mok23prl,lopez23prl,tiranov23sci}, observing subradiance is more challenging than superradiance due to the complex decay channels in most atomic systems. Subradiance has nonetheless been experimentally demonstrated in ultracold atomic \cite{guerin16prl,rui20nat,cipris21prl,ferioli21prx}, molecular \cite{takasu12prl,mcguyer15np}, and superconducting \cite{wang20prl} systems. For a perfect subradiant state, the initially excited quantum system remains in its initial state indefinitely. These decoherence-free dark states are essential for quantum information processing. The simplest perfect dark state is an antisymmetric two-atom state$(\ket{01}-\ket{10})/\sqrt{2}$, with maximum incoherent interaction, where the spontaneous emission rate approaches zero \cite{gon11prl}. However, in the vacuum, the inter-atomic interaction decays exponentially with distance. While strong, long-range interactions can be achieved in plasmonic or photonic systems \cite{mar10plsnano,pus10prb,mlynek14nc,goban15prl,ren16prb,de16acsn,ren17jcp,ren19njp}, they have strict requirements on the number and geometry of atoms, and inherent losses in these systems limit their applications in quantum information.

The study of bound states in the continuum (BIC) has emerged as a promising approach for controlling and confining optical resonant modes with remarkably high-quality factors, enabling various applications \cite{kodigala17nat,Muhammad21nl,dyakov21lpr,kang23nrp}. Of particular interest are nonradiating BIC modes that also exhibit an effective zero refractive index \cite{minkov18prl,peng20arxiv,joseph21nanopho}. Such BIC-zero-index systems can provide high-quality light confinement and facilitate long-range \cite{Ziolkowski04pre,alu07prb,Engheta13sci,maas13np,lib17np,li19prb,zhu23oe,Liedl24prx}, even all-to-all interactions \cite{ren24prapp}, owing to the infinite wavelengths inherent to these materials. These unique properties make BIC-zero-index systems a potential candidate for realizing autonomous steady-state subradiance. However, for systems with long-range dissipative interactions, the dynamical evolution of multiple particles often becomes complex, as it requires numerical integration in the exponentially growing Hilbert space, which greatly limits the understanding of phenomena such as subradiance and related research. 

In this work, we introduce a theoretical framework that addresses the dynamics of single-excitation processes in complex systems and facilitates the prediction of dynamics for long-lived multipartite entanglement, circumventing the need for intricate integral computations. Using the quantum jump operator method, we analytically derive the steady-state final state for any single-excitation initial state, including the approximate solution in the quasi-all-to-all case. The steady states exhibit W-class entanglement, with a maximally entangled state provided. As a demonstration, we compute the coupling and dynamics of emitters embedded in a dielectric photonic crystal slab with a BIC, confirming the validity of our approach. This theoretical framework elucidates a pathway toward realizing autonomous steady-state subradiance and multipartite entanglement.

This paper is organized as follows: In Sec.~II, we introduce the model, the quantum jump operator profiles, and the maximally entangled and steady state in the all-to-all case. In Sec.~III, we work out the analytical final state of any single-excitation initial state. In
Sec.~IV, we illustrate an example of emitters embedded in a high-quality photonic crystal slab possessing the bound state in the continuum to validate the efficacy of our theoretical approach. Finally, we conclude in Sec.~V.

\section{The model and perfect subradiant states within a single-excitation basis}
\subsection{The quantum jump operator profile}

Following the tracing of the environment and the application of the Born-Markov and rotating-wave approximations, the Lindblad master equation of the $n$-emitter system in a weakly coupled environment can be expressed as \cite{dung02pra,gon11prl,ren19njp}
\begin{equation}
	\frac{\partial \rho}{\partial t}=\frac{i}{\hbar}[\rho, H]+\frac{1}{2} \sum_{i, j} \gamma_{i j}\left(2 \hat{\sigma}_i \rho \hat{\sigma}_j^{\dagger}-\rho \hat{\sigma}_i^{\dagger} \hat{\sigma}_j-\hat{\sigma}_i^{\dagger} \hat{\sigma}_j \rho\right),
\end{equation}
where the raising and lowering operators of the $i$th emitter are denoted by $\hat{\sigma}_i^{\dagger}$ and $\hat{\sigma}_i$, respectively. The Hamiltonian shown in Eq.~(1) is
\begin{equation}	H=\hbar\omega_0\sum_{i}\hat{\sigma}_i^{\dagger}\hat{\sigma}_i+\sum_{i\neq j}g_{ij}\hat{\sigma}_i^{\dagger}\hat{\sigma}_j,
\end{equation}
with $g_{ij}$ being the coherent coupling between the $i$th and $j$th emitters and $\omega_0$ being the transition frequency of emitters (here we assume that the two-level emitters are identical).

When discussing collective emission dynamics, the spin operators of single emitters can be conveniently recast into the collective jump operator $\{ \hat{\mathcal{O}}_{\nu}\}$, and the Lindblad equation gives
\begin{equation}
	\frac{\partial \rho}{\partial t}=\frac{i}{\hbar}[\rho, H]+\frac{1}{2} \sum_{\nu} \Gamma_{\nu}\left(2 \hat{\mathcal{O}}_{\nu} \rho \hat{\mathcal{O}}_{\nu}^{\dagger}-\rho \hat{\mathcal{O}}_{\nu}^{\dagger} \hat{\mathcal{O}}_{\nu}-\hat{\mathcal{O}}_{\nu}^{\dagger} \hat{\mathcal{O}}_{\nu} \rho\right),
\end{equation}
with $\left\lbrace \Gamma_{\nu}\right\rbrace $ being the eigenvalues of the decoherence matrix $\boldsymbol \Gamma$, which takes the form
\begin{equation}
	\begin{split}
		&\boldsymbol \Gamma=
		\begin{pmatrix} \gamma_{11} & \gamma_{12} & \cdots & \gamma_{1n}  \\ \gamma_{21} & \gamma_{22} & \cdots & \gamma_{2n} \\ \vdots & \vdots & \ddots & \vdots \\ \gamma_{n1} & \gamma_{n2} & \cdots & \gamma_{nn} \end{pmatrix},
	\end{split}
\end{equation}
where the matrix element $\gamma_{ij}$ is the dissipative coupling between the $i$th and $j$th emitters, which can be calculated using electromagnetic Green's tensor, as shown in Appendix C, and correspondingly, we can find that the collective operators $\hat{\mathcal{O}}_{\nu}$ and $\hat{\mathcal{O}}_{\nu}^{\dagger}$ are
\begin{equation}	\hat{\mathcal{O}}_{\nu}=\sum_{i=1}^{n}\alpha_{\nu,i}\hat{\sigma}_{i}, \hat{\mathcal{O}}_{\nu}^{\dagger}=\sum_{i=1}^{n}\alpha_{\nu,i}^{*}\hat{\sigma}_{i}^{\dagger},
\end{equation}
where $\left( \alpha_{\nu,1},\alpha_{\nu,2},\cdots,\alpha_{\nu,n}\right) ^T$ is the normalized eigenvector of matrix $\boldsymbol \Gamma$, which corresponds to the eigenvalue $\Gamma_{\nu}$. Consequently, we have
\begin{equation}	\sum_{i=1}^{n}\alpha_{\nu,i}^{*}\alpha_{\mu,i}=\delta_{\nu\mu},
\end{equation}
and the matrix elements of $\boldsymbol \Gamma$ as illustrated in Eq.~(3) are
\begin{equation}
	\gamma_{ij}=\sum\limits_{\nu=1}^n \Gamma_{\nu}\alpha_{\nu,i}\alpha_{\nu,j}^*.
\end{equation}

In the language of this collective operator, a generic form of a master equation can be written as
\begin{equation}
	\frac{\partial \rho}{\partial t}=\frac{i}{\hbar}[\rho, H]+\sum_{\nu} \mathcal{L_\nu} \rho,
\end{equation}
where the $\nu$-jump Liouvillian superoperator $\mathcal{L_\nu}$ is
\begin{equation}
\mathcal{L_\nu}\left( \raisebox{-0.3em}{\scalebox{2}{$\cdot$}} \right)= \Gamma_{\nu}[ \hat{\mathcal{O}}_{\nu} \left( \raisebox{-0.3em}{\scalebox{2}{$\cdot$}} \right)  \hat{\mathcal{O}}_{\nu}^{\dagger}-\frac{1}{2}\left( \raisebox{-0.3em}{\scalebox{2}{$\cdot$}} \right) \hat{\mathcal{O}}_{\nu}^{\dagger} \hat{\mathcal{O}}_{\nu}-\frac{1}{2}\hat{\mathcal{O}}_{\nu}^{\dagger} \hat{\mathcal{O}}_{\nu} \left( \raisebox{-0.3em}{\scalebox{2}{$\cdot$}} \right)].
\end{equation}
The notation $\left( \raisebox{-0.3em}{\scalebox{2}{$\cdot$}} \right)$ represents a placeholder for an arbitrary operator or state on which the superoperator $\mathcal{L_\nu}$ acts.

\subsection{The solution of master equation}
It is convenient to convert the matrices, such as density matrix, $\rho$ to flattened vector $\tilde{\rho}$ following a column-wise ordering
\begin{equation}
	\rho=
	\begin{pmatrix} \rho_{11} & \rho_{12} \\
		\rho_{21} & \rho_{22} \end{pmatrix} \to 
	\tilde{\rho}=
	\begin{pmatrix} \rho_{11} \\ \rho_{21} \\
		\rho_{12} \\ \rho_{22} \end{pmatrix}.
\end{equation}
This process is the flattening of a matrix. For a general matrix $A$, it can be flattened as
\begin{equation}
	\begin{split}
		& A=
		\begin{pmatrix} a_{11} & a_{12} & \cdots & a_{1n}  \\ a_{21} & a_{22} & \cdots & \gamma_{2n} \\ \vdots & \vdots & \ddots & \vdots \\ a_{n1} & a_{n2} & \cdots & a_{nn} \end{pmatrix} \to 
		\tilde{A}=
		\begin{pmatrix} \boldsymbol{a}_{1}  \\ \boldsymbol{a}_{2} \\ \vdots \\ \boldsymbol{a}_{n}  \end{pmatrix},
	\end{split}
\end{equation}
where $\boldsymbol{a}_i$ is the $i$th column of $A$ with $i\in [1,n]$. In this work, we call the vector $\tilde{A}$ the flattened form of matrix $A$.

In this context, we can transfer the superoperators, as illustrated in Eq.~(8), into the multiplication of $\rho$ by a matrix for convenience. We denote the multiplication operations $A\left( \raisebox{-0.3em}{\scalebox{2}{$\cdot$}} \right) $ and $\left( \raisebox{-0.3em}{\scalebox{2}{$\cdot$}} \right)B $ as $pre(A)$ and $post(B)$, respectively. Here we assume the dimensions of matrix $A$, $B$ and $\rho$ are all $n$, and we can easily obtain that
\begin{equation}
	\widetilde{A \rho} \equiv pre(A)\tilde{\rho}=
	(\mathbf{I}_{2^n}\otimes A) \tilde{\rho},
\end{equation}
and
\begin{equation}
	\widetilde{ \rho B} \equiv post(B)\tilde{\rho}=
	(B^\top \otimes \mathbf{I}_{2^n}) \tilde{\rho},
\end{equation}
where $pre(A)$ and $post(A)$ are $(2^n)^2$-dimensional matrices and $\top$ represents the transpose of a matrix.

Therefore, the superoperator of the right-hand side of Eq.~(8) can be written as a $(2^n)^2\times (2^n)^2$ matrix
\begin{align}
	\begin{split}
		\mathcal{L} &= \frac{i}{\hbar} \left[ post(H)-pre(H)\right] \\
		&+\!\! \sum_{\nu}\Gamma_\nu \!\!\left[pre(\!\hat{\mathcal{O}}_{\nu}\!)post(\!\hat{\mathcal{O}}_{\nu}^{\dagger}\!) \!- \! pre(\!\hat{\mathcal{O}}_{\nu}^{\dagger} \hat{\mathcal{O}}_{\nu}\!)\!/2\!-\! post(\!\hat{\mathcal{O}}_{\nu}^{\dagger} \hat{\mathcal{O}}_{\nu}\!)\!/2 \right],
	\end{split}	
\end{align}
and thus the Lindblad master equation can be written in flattened form
\begin{equation}
	\frac{\partial \tilde{\rho}}{\partial t}=\mathcal{L} \tilde{\rho}.
\end{equation}

The dimension of $\mathcal{L}$ grows with $(2^n)^2$, so when $n$ is large, the numerical procedure will be very complicated. Subsequently, we will give a significantly simplified analytical method for the all-to-all case to calculate the system's steady state over an extended period.

Actually, the flattened density matrix $\tilde{\rho}(t)$ is a vector, so we rewrite this vector as a traditional Dirac form $\ket{r(t)}$. It is important to observe that the vector presented here is not a conventional pure state but a vector form of a flattened density matrix. The initial state is denoted as $\ket{r(0)}$. In this case, $\ket{r(0)}=\mathrm{flatten}\{\proj{\psi_0}\}$, where $\ket{\psi_0}$ represents the actual initial state. After the diagonalization procedure, we can obtain the eigenvectors and the eigenvalues of matrix $\mathcal{L}$. Nevertheless, these eigenvectors are not orthogonal. Subsequently, we orthogonalize these eigenvectors to obtain the orthogonalized and normalized eigenvectors $\left\lbrace \ket{l_\beta}\right\rbrace $ and their corresponding eigenvalues $\left\lbrace \epsilon_\beta\right\rbrace $, namely $\mathcal{L}=\sum_{\beta}\epsilon_\beta\proj{l_\beta}$. A diagonalization procedure can be found in Appendix B. Consequently, the state at time $t$ can be readily obtained as
\begin{equation}
	\begin{split}
		\ket{r(t)}&=e^{\mathcal{L}t}\ket{r(0)}\\
		&=\sum_{\beta}\bra{l_\beta}r(0)\rangle e^{\epsilon_\beta t}\ket{l_\beta}.
	\end{split}
\end{equation}
It is important to note that it is generally challenging to diagonalize and orthogonalize a system with a large $n$, as the dimension of the matrix to be diagonalized increases with $(2^n)^2$. However, as we will see below, it is feasible to identify an analytical solution in the all-to-all scenario.

\subsection{Maximally entangled and steady state in the all-to-all case}
In the all-to-all case, the elements of the decoherence matrix $\boldsymbol \Gamma$ are all 1. The only non-zero eigenvalue of $\boldsymbol \Gamma$ is $\Gamma_{1}=n$, and the corresponding eigenvector is $\frac{1}{\sqrt{n}}\left( 1,1,\cdots,1\right)^\top$. Therefore, the master equation depicted in Eq.~(8) can be written as
\begin{equation}
	\frac{\partial \rho}{\partial t}=\frac{i}{\hbar}[\rho, H]+\frac{n}{2} \left(2 \hat{\mathcal{O}}_{1} \rho \hat{\mathcal{O}}_{1}^{\dagger}-\rho \hat{\mathcal{O}}_{1}^{\dagger} \hat{\mathcal{O}}_{1}-\hat{\mathcal{O}}_{1}^{\dagger} \hat{\mathcal{O}}_{1} \rho\right)
\end{equation}
with 
\begin{equation}
	\hat{\mathcal{O}}_{1}=\frac{1}{\sqrt{n}}\sum_{i=1}^{n}\hat{\sigma}_{i}, \hat{\mathcal{O}}_{1}^{\dagger}=\frac{1}{\sqrt{n}}\sum_{i=1}^{n}\hat{\sigma}_{i}^{\dagger}.
\end{equation}
The matrix $\mathcal{L}$ is deterministic for a qubit system with a deterministic Hamiltonian and dissipative couplings between any two qubits. A steady state $\tilde{\rho}_{\rm s}$ for such a dissipative system requires that $\mathcal{L}\tilde{\rho}_{\rm s}=0$. The primary objective of this research is to identify non-trivial steady states that will enable us to determine the quantum state of the system after an infinitely long period. Then, we will provide a comprehensive explanation of how to utilize non-trivial steady states to obtain the exact arbitrarily long-time quantum state in the all-to-all case and the approximate long-time quantum states in the nearly all-to-all cases.

We first define the single-excitation basis $\{ \ket{\boldsymbol{1}}=\ket{00\cdots1}, \ket{\boldsymbol{2}}=\ket{00\cdots10}, \cdots, \ket{\boldsymbol{n}}=\ket{10\cdots00}\}$. The number $\boldsymbol{i}$ in $\ket{\boldsymbol{i}}$ corresponds to the $i$th qubit is excited and the others are in their ground states, and $\ket{\boldsymbol{0}}$ is the vacuum state. The maximal multipartite entanglement state within the single-excitation regime is the so-called W-state (or single-excitation Dicke state), which has the form 
\begin{equation}
    \ket{W_n}=\frac{1}{\sqrt{n}}\left( \ket{\boldsymbol{1}}+\ket{\boldsymbol{2}}+\cdots \ket{\boldsymbol{n}}\right).
\end{equation}
Here, we use negativity, an easily computable bipartite entanglement measure, to measure the bipartite and multipartite entanglement. For bipartite mixed state $\rho_{AB}$, its negativity is defined as $N(\rho_{AB})=||\rho^{T_A}_{AB}||_1-1=\sum_k|\lambda_k|-1=2\sum_k|\lambda_k^{neg}|$, where $\lambda_ks$ and $\lambda_k^{neg}s$ are the eigenvalues and negative eigenvalues of partial transposition matrix $\rho^{T_A}_{AB}$, respectively \cite{ver02pra}. Bennett \emph{et al} pointed out that a state has genuine $n$-partite correlations if it is nonproduct in every bipartite cut \cite{ben11pra}. Therefore, via the bipartite entanglement negativity, we can define a measure for genuine multipartite quantum entanglement
\begin{equation}\label{3}
	N_n(\rho_{n})=[N(\rho_{i_1|\overline{i}_1}) \cdot N(\rho_{i_2|\overline{i}_2}) \cdots N(\rho_{i_n|\overline{i}_n})]^{1/n},
\end{equation}
where $\rho_{n}$ is a $n$-qubit state, and the bipartite negativity $N(\rho_{i_k|\overline{i}_k})$ characterizes the quantum correlation between the $k$th qubit $i_k$ and the rest of the system $\overline{i}_k$. Therefore, the multipartite quantum correlation $N_n(\rho_{n})$ equals zero when the multipartite mixed state is biseparable in any partition. This measure has been extensively employed in numerous many-body systems \cite{bayat17prl,su22pra,su24prb}.

The only negative eigenvalue of the partial transpose of the W state is $\lambda^{neg}=-{\sqrt{n-1}}/{n}$, therefore, the multipartite negativity $N_n$ can be calculated as $N_n=2\sqrt{n-1}/n$. In reality, not only the normal W state with identical coefficients, for a state of equal probability superposition of $\ket{\boldsymbol{i}}$, $\sum_{i}c_i \ket{\boldsymbol{i}}$ with $|c_i|=1/\sqrt{n}$, the multipartite quantum entanglement $N_n$ is equal to the W state. It can be determined that if the system is prepared at the superposition of antisymmetric state $\ket{\boldsymbol{i}}-\ket{\boldsymbol{j}}$ initially, the system will always stay at this state, which means that $\ket{\boldsymbol{i}}-\ket{\boldsymbol{j}}$ along with their superpositions are steady states of the all-to-all system. Consequently, we can conclude that, in the all-to-all $n$-qubit system, the maximally entangled and perfectly subradiant state can be described as
\begin{equation}
\begin{split}
    &\ket{\psi_{\rm s}}\!=\!\sum_{i\neq j} R_{ij}(\ket{\boldsymbol{i}}-\ket{\boldsymbol{j}}) \\
    \text{s.t.}\  &\ket{\psi_{\rm s}}\!=\!\sum_{i=1}^n c_{i}\ket{\boldsymbol{i}},  \ |c_i|^2 \!=\! \frac{1}{n} \  \forall i \in \! \{1, 2, \ldots, n\},
\end{split}
\end{equation}
where $R_{ij}$s are random complex numbers. Therefore, if the initial state is prepared as in Eq.~(21), it is a random superposition of antisymmetric steady states $\ket{\boldsymbol{i}}-\ket{\boldsymbol{j}}$, and the modulus of the coefficients of each single-excitation component $\ket{\boldsymbol{i}}$ is equal, the system will remain in this $n$-qubit maximally entangled state. For an even-qubit system, such as a four-qubit system, a steady and maximally entangled state is given by 
\begin{equation}
    \ket{\psi_{\rm s}^4}=\frac{1}{2}(\ket{\boldsymbol{1}}-\ket{\boldsymbol{2}}+\ket{\boldsymbol{3}}-\ket{\boldsymbol{4}}).
\end{equation} 
For an odd-qubit system, like a three-qubit system, a perfectly subradiant and maximally entangled state takes the form
\begin{equation}
    \ket{\psi_{\rm s}^3}=\frac{1}{2\sqrt{3}}[(1+\sqrt{3}i)(\ket{\boldsymbol{1}}-\ket{\boldsymbol{2}})+(1-\sqrt{3}i)(\ket{\boldsymbol{1}}-\ket{\boldsymbol{3}})].
\end{equation} 
In the rest of this section and the next section, we take a six-qubit system as an example to illustrate the coincidence between the theoretical and numerical results. Panels (a)-(c) of Fig.~1 illustrate the dynamics of multipartite entanglement $N_6$ and bipartite entanglement $N_{half}$ (the entanglement between one half of the system and the rest) for three initial states
\begin{equation}
    \begin{split}
        \ket{r_1}&=\frac{1}{\sqrt{6}}(\ket{\boldsymbol{1}}-\ket{\boldsymbol{2}}+\ket{\boldsymbol{3}}-\ket{\boldsymbol{4}}+\ket{\boldsymbol{5}}-\ket{\boldsymbol{6}}), \\
        \ket{r_2}&=\frac{1}{\sqrt{258}}[8(\ket{\boldsymbol{1}}-\ket{\boldsymbol{2}})+\ket{\boldsymbol{3}}-\ket{\boldsymbol{4}}+8(\ket{\boldsymbol{5}}-\ket{\boldsymbol{6}})], \\
        \ket{r_3}&=\frac{1}{\sqrt{2}}(\ket{\boldsymbol{1}}-\ket{\boldsymbol{4}}),
    \end{split}
\end{equation}
respectively. The numerical calculation results (symbols) and theoretical predictions (dashed lines calculated using initial states) are consistent, and the states remain unchanged despite having completely different entanglement distributions. Obviously, all three states are antisymmetric states or their linear superpositions, so they will always stay at the initial states, that is, perfect subradiant states. However, only $\ket{r_1}$ fully satisfies the conditions of Eq.~(21), namely the maximally entangled steady state; $\ket{r_2}$ is a partially multipartite entangled state and $\ket{r_3}$ has only bipartite entanglement. 

\begin{figure}
	\includegraphics[width=0.5\textwidth]{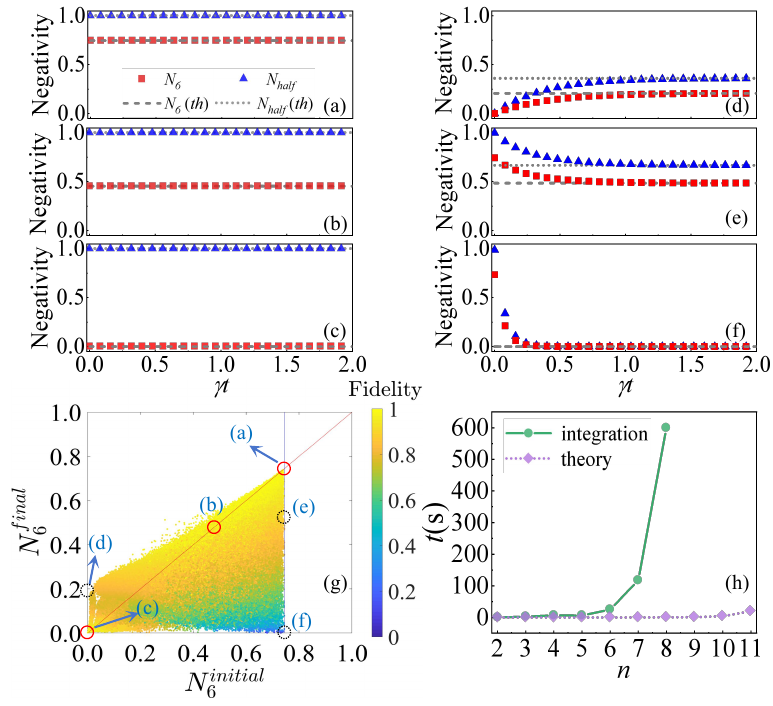}
	\caption{(Color online) Evolution of multipartite quantum correlation $N_6$ (red squares) and bipartite entanglement $N_{half}$ (blue triangles) with six typical initial states, $\ket{r_1}\sim \ket{r_6}$ shown in the main text correspond to panel (a)$\sim$(f), respectively. Gray dashed and dotted lines represent $N_6(th)$ and $N_{half}(th)$ calculated by the final states that are predicted by our theory. (g) Relationship between the multipartite entanglement of the totally random single-excitation states and their final-state result. The color of the scattered points represents the fidelity between the initial and final states. (h) A comparison of the calculation time between the numerical and our approach with the same initial state $\ket{\boldsymbol{1}}$.}
\end{figure}

% The comparison between the initial and final entanglement of 200 random steady defined in Eq.~(8) further illustrates this conclusion, as shown in panel (d).

\section{The analytical final state of any single-excitation initial state}
Among the eigenvalues of the matrix $\mathcal{L}$ for a $n$-qubit single-excitation system, only a few possess a zero real part (steady-state basis), while the remaining eigenvalues are negative (exponentially decaying basis). Consequently, once the eigenvector corresponding to the eigenvalues with a zero part is identified, the long-term steady state can be determined. The zero-eigenvalue basis of matrix $\mathcal{L}$ is the flattened vectors of
\begin{equation}
	a_k \! = \!
	\begin{cases} 
		\frac{1}{2}(\ket{\boldsymbol{1}}-\ket{\boldsymbol{i}})\otimes(\bra{\boldsymbol{1}}-\bra{\boldsymbol{i}}), & \text{if } k \in [1, n-1],  \\
		\frac{1}{2}(\ket{\boldsymbol{1}}-\ket{\boldsymbol{i}})\otimes(\bra{\boldsymbol{1}}-\bra{\boldsymbol{j}}), & \text{if } k \in [n, (n-1)^2],
	\end{cases}
\end{equation}
and the purely imaginary-eigenvalue basis (the eigenvalues are $\pm i\omega_0$) is the flattened vectors of 
\begin{equation}
	b_k \!=\!
	\begin{cases} 
		\frac{1}{\sqrt{2}}(\ket{\boldsymbol{i}}\bra{\boldsymbol{0}}-\ket{\boldsymbol{1}}\bra{\boldsymbol{0}}), & \text{if } k \in [1, n-1],  \, \epsilon_k = -i\omega_0, \\
		\frac{1}{\sqrt{2}}(\ket{\boldsymbol{0}}\bra{\boldsymbol{i}}-\ket{\boldsymbol{0}}\bra{\boldsymbol{1}}), & \text{if } k \in [n, 2(n-1)], \,  \epsilon_k = i\omega_0,
	\end{cases}
\end{equation}
where $\boldsymbol{i}, \boldsymbol{j}\in [\boldsymbol{2}, \boldsymbol{n}]$ and $ \boldsymbol{i} \neq \boldsymbol{j}$, and $\epsilon_k$ is the corresponding eigenvalue of $b_k$. Obviously, $\{\tilde{a}_k\} $ are not orthogonal and neither are $\{\tilde{b}_k\} $. To obtain the final state, we need to orthogonalize and normalize $\{\tilde{a}_k\} $ and $\{\tilde{b}_k\} $ (note that $a_k$ and $b_k$ are orthogonal) as $\{\tilde{A}_k\} $ and $\{\tilde{B}_k\} $, respectively. The detailed orthogonalization process is shown in Appendix B, where we also give the specific process for the three-qubit case.

For a single-excitation initial state $\ket{r(0)}$, its corresponding state at a large time $t$ can be readily obtained as
\begin{equation}
	\begin{split}
		\ket{r(t)}&=\sum_{k=1}^{(n-1)^2}\bra{\tilde{A}_k}r(0)\rangle \ket{\tilde{A}_k}+\sum_{k=1}^{n-1}\bra{\tilde{B}_k}r(0)\rangle e^{-i\omega_0 t}\ket{\tilde{B}_k}\\
		&+\sum_{k=n}^{2(n-1)}\bra{\tilde{B}_k}r(0)\rangle e^{i\omega_0 t}\ket{\tilde{B}_k}+c_0\ket{\tilde{v}},
	\end{split}
\end{equation}
where the vector $\ket{\tilde{v}}$ represents the flattened form of the vacuum state $\ket{\boldsymbol{0}}\bra{\boldsymbol{0}}$, while the term $c_0\ket{\tilde{v}}$ is included to ensure that the trace of the final state equals 1. Except for the eigenvectors $\tilde{A}_k$ and $\tilde{B}_k$ corresponding to Eqs.~(25) and (26), the eigenvalues of the remaining eigenvectors of matrix $\mathcal{L}$ all contain negative eigenvalues. Therefore, the dynamical evolution terms corresponding to these eigenvalues will decay exponentially as time increases, and thus will not be included in the final quantum state with infinite time shown in Eq.~(27). 

Panels (d)-(f) of Fig.~1 are the multipartite and bipartite entanglement dynamic $N_6$ and $N_{half}$ of three typical six-qubit initial states 
\begin{equation}
    \begin{split}
        \ket{r_4}&=\ket{\boldsymbol{1}}, \\
        \ket{r_5}&=\frac{1}{\sqrt{6}}(\ket{\boldsymbol{1}}+\ket{\boldsymbol{2}}+\ket{\boldsymbol{3}}-\ket{\boldsymbol{4}}-\ket{\boldsymbol{5}}+\ket{\boldsymbol{6}}), \\
        \ket{r_6}&=\ket{W_6}.
    \end{split}
\end{equation}
The gray dashed and dotted lines in panels (d)-(f) of Fig.~1 represent $N_6(th)$ and $N_{half}(th)$ calculated by the final states that are theoretically predicted by Eq.~(27). The coincidence of the theoretical and numerical results at a large time verifies that, even if the initial state is not a steady state, our theory can accurately predict its final state, whether it is the generation [panel (d)], reduction [panel (e)] or disappearance [panel (f)] of entanglement. Specifically, if the system is initially prepared in the maximally entangled W state [see panel (f)], it will decay rapidly and evolve into a product state. Nevertheless, if the system is to be maintained in the maximally entangled state, the initial state must be prepared in a form shown in panel (a).

Figure 1(g) shows the relationship between the multipartite entanglement of totally random single-excitation states and their final-state result. The color of the scattered points in the panel represents the fidelity \cite{nielsen00book} between the initial and final states. The points on the diagonal correspond to the steady-state subradiance indicated by the first line of Eq.~(21), where the three points marked by solid red circles represent Fig.~1(a)-1(c), and their fidelity is 1. The other three dotted black circles correspond to the three situations shown in Fig.~1(d)-1(f). Panel (h) compares the calculation time between the numerical and our approach, demonstrating our theory's efficiency.  Therefore, we can immediately get the exact form of any initial state's final steady state. More importantly, we can easily find out what initial state the system is prepared to maintain its initial quantum state unchanged. This has a far-reaching significance for the production and storage of quantum resources.
\begin{figure}
\includegraphics[width=0.5\textwidth]{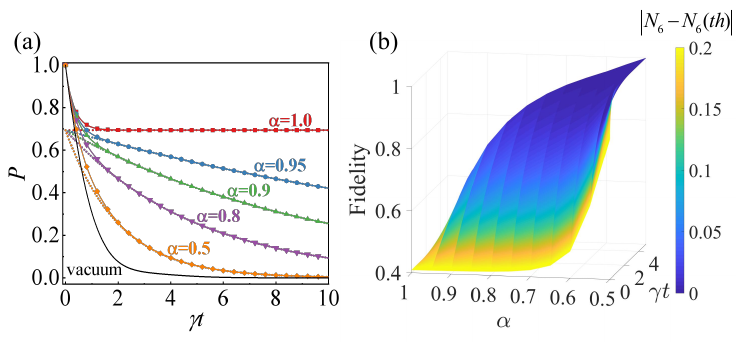}
\caption{(Color online) (a) The numerical calculation of probability $P$ of the system remains in the initial state $\ket{r(0)}=\ket{\boldsymbol{1}}$ (marks) and our theoretical prediction (dashed lines) under different $\alpha$s. (b) The fidelity between the numerical and theoretical states as the function of time and $\alpha$. The color represents the difference of $N_6$ between the numerical and theoretical states.}
\end{figure}

Suppose the incoherent coupling between any two qubits in a $n$-qubit system is a value $\alpha$ that is close to 1, the steady basis shown in Eqs.~(25) and (26) have the corresponding eigenvalues with near-zero real parts $\alpha-1$ and $\alpha-1\pm i\omega_0$, respectively. In that case, the state at a large time as shown in Eq.~(27) can be approximately written as
\begin{equation}
	\begin{split}
		\ket{r(t)}&\approx c_0\ket{\tilde{v}}+\sum_{k=1}^{(n-1)^2}\bra{\tilde{A}_k}r(0)\rangle e^{-(1-\alpha)t} \ket{\tilde{A}_k}\\
		&+\sum_{k=1}^{n-1}\bra{\tilde{B}_k}r(0)\rangle e^{[-(1-\alpha)-i\omega_0 ]t}\ket{\tilde{B}_k}\\
		&+\sum_{k=n}^{2(n-1)}\bra{\tilde{B}_k}r(0)\rangle e^{[-(1-\alpha)+i\omega_0 ]t}\ket{\tilde{B}_k}.
	\end{split}
\end{equation}
All the other eigenvalues have finite and negative real parts, so their corresponding dynamics will decay exponentially, thus they are not included in the final states. The detailed analysis can be found in Appendix A and B. In Fig.~2(a), we plot the probability $P=\bra{r(0)}\rho(t)\ket{r(0)}$ under different $\alpha$s (shown by different symbols), which is the probability of the system remains in the initial state $\ket{r(0)}=\ket{\boldsymbol{1}}$ of a six-qubit initial state. The dotted lines are the corresponding theoretical results predicted by Eq.~(29). As time increases, theoretical and numerical populations with very different initial values quickly approach, and they almost coincide in the end, regardless of the value of $\alpha$. To further demonstrate the efficacy of the theory under different $\alpha$s, we plot the dependence of fidelity between the theoretical and numerical results on $\alpha$ and time $t$ in Fig.~2(b). The colors in the panel represent the difference between the theoretically and numerically calculated multipartite entanglement $N_6$. When time is large, the numerical value of entanglement and the quantum state predicted by our theory is very close to the result of the numerical calculation. This result confirms the validity of Eq.~(27), which can accurately predict the evolution of quantum states over a long time even if the system is not a perfect all-to-all situation.

In the above study, we did not consider the effect of coherent dipole-dipole coupling in the system. In fact, the effect of coherent coupling on the emission dynamics of the system is very limited, and the relevant discussions can be found in the next section and Appendix A.

\begin{figure}
	\includegraphics[width=0.5\textwidth]{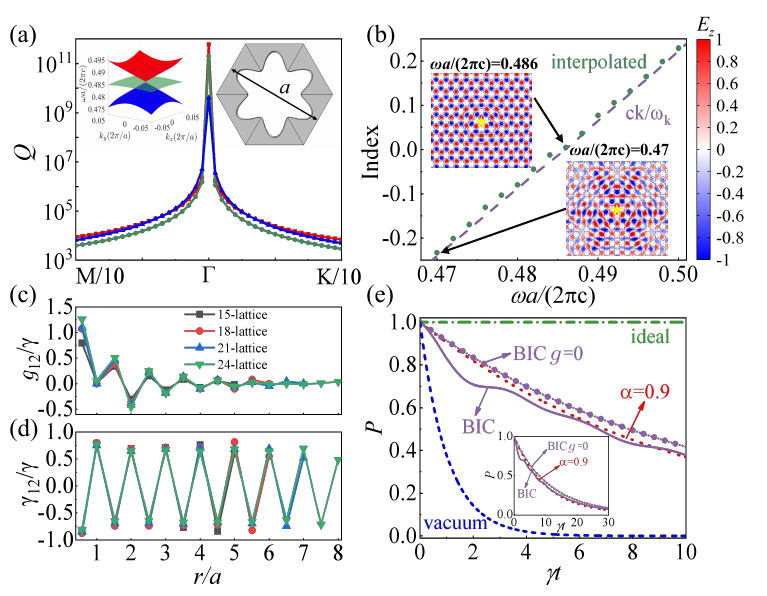}
	\caption{(Color online) (a) Divergent quality factors of Dirac cone and flat band (energy band is shown in the left inset) of a daisy photonic crystal slab as shown in the right inset. (b) Effective index calculated from the energy band (dashed purple line) and transmission of electromagnetic field (dotted green symbols). Two insets show the field distribution emitted from a single excited emitter (yellow star) embedded in the slab near the $\Gamma$ point (top left inset) and far away from $\Gamma$ (bottom right inset). (c) and (d) Calculated coherent and incoherent couplings of emitters placed at high symmetry points for different slab sizes. (e) Subradiant state evolutions of six emitters under several scenarios: the ideal all-to-all case (dash-dotted green line), evolution predicted by Eq.~(29) with $\alpha=0.9$ (dotted red line), evolution calculated in the BIC mode with (solid purple line) and without coherent coupling (dotted purple line), and the situation in vacuum (dashed blue line).}
\end{figure}

\section{Subradiance in a bound state in the continuum}
Here, we use the emitters embedded in an all-dielectric daisy photonic crystal slab [as shown in the right inset of Fig.~3(a)] with zero-index mode which is also symmetry-protected BIC \cite{minkov18prl} to realize the low-loss and all-to-all interactions. As shown in the energy band around $\Gamma$ point that is presented in the left inset of Fig.~3(a), there is a clear Dirac cone and a flat band at $\Gamma$ point, and the Q-factor of all three bands diverges to infinity. In Fig.~3(b), we present the effective index derived through the calculation and interpolation of electric field in high-symmetry positions (dotted green line) $E_z(r+m\sqrt3a)=E_z(r)e^{in_ek_0m\sqrt3a}$ \cite{minkov18prl}, where $k_0 = \omega/c$ and $m$ takes integer values, and the dashed line is derived directly from the energy band near $\Gamma$ point, $n_e=ck/\omega_k$. Both simulation results confirm an effective near-zero index mode near the frequency $\omega a/(2\pi c)=0.486$ [as shown in Fig.~3(b)]. The two insets in Fig.~3(b) show the normalized spatial field diagram of placing an excited dipole emitter at a high-symmetry point (marked by yellow stars in two insets) when the transition frequency of emitter is near the $\Gamma$ point $[\omega a/(2\pi c)=0.486]$ or far away from the $\Gamma$ point $[\omega a/(2\pi c)=0.47]$, respectively. As can be seen from two insets, at the $\Gamma$ point, the near-zero index causes the resonant emitter to produce a highly symmetric and almost non-attenuated field extension in space. However, when the frequency deviates from the $\Gamma$ point, the spatial field distribution of the emitter is nonuniform. More details about the photonic crystal slab as well as the field distribution diagrams can be found in Appendix C. The electric field patterns are calculated with a lattice constant $a=800~\rm nm$. This means that in this high-quality BIC system, near-zero-index effects can be observed on a large spatial scale, which suggests that it may be possible to achieve long-range, unattenuated interactions on a large spatial scale.

Figure 3(c) and 3(d) are the calculated normalized pairwise coherent and incoherent couplings of two emitters, $g_{12}/\gamma$ and $\gamma_{12}/\gamma$, placed at high-symmetry points for different slab sizes, where one emitter is fixed in the center of the slab sample that possesses the spontaneous decay rate $\gamma$, and the other has a distance $r$ with it. The coherent coupling between two emitters can be denoted by 
\begin{equation}
	\begin{split}
     g_{12}=\frac{\omega_0^2}{\varepsilon_0\hbar c^2}{\rm Re}[\vec\mu_1^\ast\cdot \overset{\leftrightarrow}{G}(\vec r_1,\vec r_2,\omega)\cdot\vec\mu_2],
    \end{split}
\end{equation}
while the incoherent coupling is
\begin{equation}
	\begin{split}
     \gamma_{12}=\frac{2\omega_0^2}{\varepsilon_0\hbar c^2}{\rm Im}[\vec\mu_1^\ast\cdot \overset{\leftrightarrow}{G}(\vec r_1,\vec r_2,\omega)\cdot\vec\mu_2].
    \end{split}
\end{equation}
The Green tensor $\overset{\leftrightarrow}{G}(\vec r_1,\vec r_2,\omega)$ showed in Eqs.~(30) and (31) represents the electric field at position $\vec r_1$ due to a source located at $\vec r_2$ with frequency $\omega$, and the distance between them is $r=|\vec r_1-\vec r_2|$. In this work, the dipole moments are all directed along the $z$ axis and have the same electric dipole moment. Therefore, the Green's function can be obtained by solving for the electric field:
\begin{equation}
	\begin{split}
     \vec\mu_1^\ast\cdot \overset{\leftrightarrow}{G}(\vec r_1,\vec r_2,\omega)\cdot\vec\mu_2=-E_z(\vec r_1)_{\vec r_2},
    \end{split}
\end{equation}
where $E_z(\vec r_1)_{\vec r_2}$ is the $z$ component of the electric field at $\vec r_1$ emitted from the emitter located at $\vec r_2$.
The detailed arrangement of emitters is shown in Appendix C.

According to Fig.~3, the coherent couplings $g_{12}/\gamma$ decrease rapidly with the emitter-emitter distance, and almost turns to zero when the distance between two emitters are large. Nevertheless, the incoherent couplings $\gamma_{12}/\gamma$ are relatively uniform, which is also a consequence of the zero index. Figure 3(e) presents the population of the maximally entangled and steady state $\ket{r(0)}=\frac{1}{\sqrt{6}}(\ket{\boldsymbol{1}}-\ket{\boldsymbol{2}}+\ket{\boldsymbol{3}}-\ket{\boldsymbol{4}}+\ket{\boldsymbol{5}}-\ket{\boldsymbol{6}})$ under several scenarios. The solid purple line represents the numerical result in the BIC system as shown in Fig.~3(a) (the emitter arrangement is shown in Appendix C), and the dashed blue line is the result for emitters in a vacuum. It can be seen that although the state evolution in the photonic crystal slab has not yet reached the ideal situation (the all-to-all case, denoted by horizontal dash-dotted green line), the rate of state decay is much lower than that in a vacuum, which is definite evidence of subradiance. 

The oscillation at the beginning of the evolution originates from the coherent coupling of the nearest neighbors, and its influence on the evolution of the entire state is very limited. This can be seen from the fact that the curves of the dynamical evolution are very close when the coherent coupling is taken as zero (the dotted purple line, denoted by BIC $g=0$) during the evolution in the BIC case. The dotted red line is the theoretical result predicted by Eq.~(29) with $\alpha=0.9$, which is very close to the calculated dynamic evolution curve in the real BIC system, even over a longer period (see inset). That confirms that our theory can predict the dynamical evolution of quantum states on a longer time scale (see inset).

\section{Summary}
In summary, we have developed successfully a theoretical framework that is able to reveal quite efficiently the dynamics of single-excitation processes in all-to-all systems, facilitating the prediction of long-lived multipartite entanglement dynamics without intricate integral computations. Utilizing the quantum jump operator method, we have analytically derived the steady-state final state for any single-excitation initial state, including an approximate solution for the quasi-all-to-all coupling scenario. Our results demonstrate unambiguously that the system's multipartite entanglement, characterized by the negativity measure, remains robust over time, with the steady states exhibiting W-class entanglement. As an illustrative example, we have computed the coupling and dynamics of emitters within a photonic crystal slab possessing an ultra-high quality BIC, thereby validating the efficacy of our theoretical approach. This work not only elucidates a pathway toward realizing autonomous steady-state subradiance and multipartite entanglement in atomic systems but also has significant implications for the production and storage of quantum resources, potentially enabling more efficient quantum information processing and quantum communication protocols.

\begin{acknowledgments}
	This work was supported by NSF-China (Grant Nos. 11904078 and 11575051), GRF (Grant Nos. 17310622 and 17303023) of Hong Kong, Hebei NSF (Grant Nos. A2021205020 and A2019205266), and Hebei 333 Talent Project (B20231005). JR was also funded by the project of the China Postdoctoral Science Foundation (Grant No. 2020M670683).
\end{acknowledgments}

\appendix

\section{Two-qubit case and the role of coherent coupling in emission dynamics}
In this Appendix, we first present the details about the simplest two-qubit system with the maximal incoherent coupling using the quantum jump operator method, and then we will discuss the dependence of coherent coupling on the emission dynamics. The Hamiltonian of the two-qubit system has the form
\begin{equation}    
H=\hbar\omega_0(\hat{\sigma}_1^{\dagger}\hat{\sigma}_1+\hat{\sigma}_2^{\dagger}\hat{\sigma}_2)+g(\hat{\sigma}_1^{\dagger}\hat{\sigma}_2+\hat{\sigma}_2^{\dagger}\hat{\sigma}_1),
\end{equation}
with $g$ being the coherent coupling between two qubits, and the master equation can be written as
\begin{equation}
    \begin{split}
        \frac{\partial \rho}{\partial t}&=\frac{i}{\hbar}[\rho, H]+2 \hat{\mathcal{O}}_{1} \rho \hat{\mathcal{O}}_{1}^{\dagger}-\rho \hat{\mathcal{O}}_{1}^{\dagger} \hat{\mathcal{O}}_{1}-\hat{\mathcal{O}}_{1}^{\dagger} \hat{\mathcal{O}}_{1} \rho\\
        &=\mathcal{L}\rho
    \end{split}
\end{equation}
with 
\begin{equation}
	\hat{\mathcal{O}}_{1}=\frac{1}{\sqrt{2}}(\hat{\sigma}_{1}+\hat{\sigma}_{2}), \hat{\mathcal{O}}_{1}^{\dagger}=\frac{1}{\sqrt{2}}(\hat{\sigma}_{1}^{\dagger}+\hat{\sigma}_{2}^{\dagger}).
\end{equation}

The eigenvalues of matrix $\mathcal{L}$ of such a two-qubit system can be easily obtained as
\begin{equation}
	\begin{split}
		&\epsilon_1=0, \epsilon_2=0, \epsilon_3=-i(\omega_0-g), \epsilon_4=i(\omega_0-g),\\ 
		&\epsilon_5=-1-2ig,   \epsilon_6=-1+2ig, \epsilon_7=\epsilon_8=-2, \\
		&\epsilon_9=\epsilon_{10}=-1-i(\omega_0+g),
		\epsilon_{11}=\epsilon_{12}=-1+i(\omega_0+g),\\ &\epsilon_{13}=-1-2i\omega_0,\epsilon_{14}=-1+2i\omega_0,\\
		&\epsilon_{15}=-2-i(\omega_0-g),\epsilon_{16}=-2+i(\omega_0-g).
	\end{split}
\end{equation}
According to the discussion in the main text, if the eigenvalue $\epsilon_\beta$ is equal to zero, such as $\epsilon_1$ and $\epsilon_2$, their corresponding dynamic terms $\bra{l_\beta}r(0)\rangle e^{\epsilon_\beta t}\ket{l_\beta}$ in $\ket{r(t)}$ will not vary with time $t$, we call these terms as the steady terms. Otherwise, if the eigenvalues of $\mathcal{L}$ contain a finite negative real part, such as $\epsilon_5\sim \epsilon_{16}$, their corresponding dynamic terms $\bra{l_\beta}r(0)\rangle e^{\epsilon_\beta t}\ket{l_\beta}$ turn to zero rapidly with the increase of time. Therefore, the contribution of these terms can be omitted for analyzing the quantum state for a long time, and we call these dynamic terms decay terms.  As for $\epsilon_3$ and $\epsilon_4$, they are pure imaginary numbers, so we call their contribution to the dynamics oscillatory terms. Although these terms cause oscillations in the quantum state, they do not change the quantum correlations of the system. Therefore, in order to find the final analytical steady state of the system evolution, we only need to consider the four dynamic evolutions corresponding to eigenvalues $\epsilon_1\sim \epsilon_4$.

The normalized eigenvectors correspond to $\epsilon_1$ $\sim$ $\epsilon_4$ are
\begin{equation}
	\begin{split}
		\ket{\overline{l}_1}&=\frac{1}{2}(0, 0, 0, 0, 0, 1, -1, 0, 0, -1, 1, 0, 0, 0, 0, 0)^\top,\\
		\ket{\overline{l}_2}&=(1, 0, 0, 0, 0, 0, 0, 0, 0, 0, 0, 0, 0, 0, 0, 0)^\top,\\
		\ket{\overline{l}_3}&=\frac{1}{\sqrt{2}}(0, -1, 1, 0, 0, 0, 0, 0, 0, 0, 0, 0, 0, 0, 0, 0)^\top,\\
		\ket{\overline{l}_4}&=\frac{1}{\sqrt{2}}(0, 0, 0, 0, -1, 0, 0, 0, 1, 0, 0, 0, 0, 0, 0, 0)^\top.
	\end{split}
\end{equation}

Obviously, $\ket{\overline{l}_2}$ corresponds to the trivial vacuum state, and these eigenvectors are naturally orthogonal. According to the notation in the main text, the corresponding matrices form of the nontrivial steady states are
\begin{equation}
	\begin{split}
		&A_1 =\frac{1}{2}(\ket{\boldsymbol{1}}-\ket{\boldsymbol{2}})\otimes(\bra{\boldsymbol{1}}-\bra{\boldsymbol{2}}), \\
		&B_1 =\frac{1}{\sqrt{2}}(-\ket{\boldsymbol{1}}\bra{\boldsymbol{0}}+\ket{\boldsymbol{2}}\bra{\boldsymbol{0}}),\\
		&B_2 =\frac{1}{\sqrt{2}}(-\ket{\boldsymbol{0}}\bra{\boldsymbol{1}}+\ket{\boldsymbol{0}}\bra{\boldsymbol{2}}),
	\end{split}	
\end{equation}

In this work, we consider that the coupling between emitters and light fields is weak, that is, the coupling strength $g$ is much smaller than the emission rate and transition frequency of emitters. Therefore, for simplicity, we write $\omega_0-g$ in $\epsilon_3$ and $\epsilon_4$ as $\omega_0$ in the following part. Note that if we need to consider the influence of $g$, we only need to change $\omega_0$ back to $\omega_0-g$. That means the coherent coupling only impacts the oscillation in the dynamics. For a long time, the state can be readily written as
\begin{equation}
	\begin{split}	\ket{r(t\to\infty)}&=c_0\ket{\tilde{v}}+\bra{\tilde{A}_1}r(0)\rangle \ket{\tilde{A}_1}\\
		&+\bra{\tilde{B}_1}r(0)\rangle e^{-i\omega_0 t}\ket{\tilde{B}_1}\\
		&+\bra{\tilde{B}_2}r(0)\rangle e^{i\omega_0 t}\ket{\tilde{B}_2}
	\end{split}
\end{equation}
with $c_0=1-\bra{\overline{l}_1}r(0) \rangle$ to ensure the unit trace of the density matrix.
Note that due to the Hermitian property of $\ket{r}$, we must have $\bra{\overline{l}_3}r\rangle=\bra{\overline{l}_4}r\rangle^*$. According to Eq.~(21), the final steady state is very dependent on the choice of the initial state $\ket{r(0)}$. Obviously, $\ket{\overline{l}_2}$ is the trivial ground state, and $\ket{\overline{l}_3}$ and $\ket{\overline{l}_4}$ represent the coherence between single-excitation state and ground state, which do not affect the evolution of the diagonal elements of the density matrix. 

Now we take some typical initial states to show the exact form of their corresponding final states. First, if the initial state $\ket{r(0)}=\ket{\overline{l}_1}$, which is the normalized steady state and corresponds to the maximally entangled state $\frac{1}{\sqrt{2}}(\ket{eg}-\ket{ge})$. Under this initial state, $\bra{\overline{l}_1}r(0) \rangle=1$, and the two-qubit system will remain forever in this maximally entangled initial state. Another typical initial state is $\ket{eg}$, namely one of the emitters is excited and the other is on its ground state, which is a more easily prepared initial state. In this case, we have $\bra{\overline{l}_1}r(0) \rangle=\frac{1}{2}$ and $\bra{\overline{l}_3}r(0) \rangle=\bra{\overline{l}_4}r(0) \rangle=0$, and thus we can obtain the final steady state as $\frac{1}{2}\ket{\overline{l}_1}+\frac{1}{2}\ket{\overline{l}_2}$, its matrix form is
\begin{equation}
	\rho_{\rm f}=\frac{1}{4}(2\ket{\boldsymbol{0}}\bra{\boldsymbol{0}}+\ket{\boldsymbol{1}}\bra{\boldsymbol{1}}+\ket{\boldsymbol{2}}\bra{\boldsymbol{2}}-\ket{\boldsymbol{1}}\bra{\boldsymbol{2}}-\ket{\boldsymbol{2}}\bra{\boldsymbol{1}}),
\end{equation}
with the two-qubit concurrence being $\frac{1}{2}$, and this is another perfect subradiant state.

%For another initial state $\frac{1}{\sqrt{2}}(\ket{gg}+\ket{eg})$, we have $\bra{\overline{l}_1}r(0) \rangle=\frac{1}{4}$ and $\bra{\overline{l}_3}r(0) \rangle=\bra{\overline{l}_4}r(0) \rangle=\frac{1}{2\sqrt{2}}$, thus the final state is $\frac{1}{4}\ket{\overline{l}_1}+\frac{3}{4}\ket{\overline{l}_2}+\frac{1}{2\sqrt{2}}e^{-i\omega_0t}\ket{\overline{l}_3}+\frac{1}{2\sqrt{2}}e^{i\omega_0t}\ket{\overline{l}_4}$, and the corresponding density matrix is
% \begin{equation}
% 	\begin{split}
% 		\rho_{\rm f}&=\frac{1}{8}(6\ket{\boldsymbol{0}}\bra{\boldsymbol{0}}+\ket{\boldsymbol{1}}\bra{\boldsymbol{1}}+\ket{\boldsymbol{2}}\bra{\boldsymbol{2}}-\ket{\boldsymbol{1}}\bra{\boldsymbol{2}}-\ket{\boldsymbol{2}}\bra{\boldsymbol{1}})\\
% 		&+\frac{1}{4}e^{-i\omega_0t}(-\ket{\boldsymbol{1}}\bra{\boldsymbol{0}}+\ket{\boldsymbol{2}}\bra{\boldsymbol{0}})\\
% 		&+\frac{1}{4}e^{i\omega_0t}(-\ket{\boldsymbol{0}}\bra{\boldsymbol{1}}+\ket{\boldsymbol{0}}\bra{\boldsymbol{2}}),
% 	\end{split}	
% \end{equation}
% and the two-qubit concurrence is $\frac{1}{4}$. 

For another special state, $\frac{1}{2}(\ket{e}+\ket{g})\otimes(\ket{e}-\ket{g})$, we find that $\bra{\overline{l}_1}r(0) \rangle=\frac{1}{2}$ (thus $c_0=\frac{1}{2}$), and $\bra{\overline{l}_3}r(0) \rangle=\bra{\overline{l}_4}r(0) \rangle=\frac{1}{2\sqrt{2}}$, thus the final state is $\frac{1}{2}\ket{\overline{l}_1}+\frac{1}{2}\ket{\overline{l}_2}+\frac{1}{2\sqrt{2}}e^{-i\omega_0t}\ket{\overline{l}_3}+\frac{1}{2\sqrt{2}}e^{i\omega_0t}\ket{\overline{l}_4}$, and the final state can be written as the following matrix form 
\begin{equation}
	\begin{split}
		\rho_{\rm f}&=\frac{1}{4}(2\ket{\boldsymbol{0}}\bra{\boldsymbol{0}}+\ket{\boldsymbol{1}}\bra{\boldsymbol{1}}+\ket{\boldsymbol{2}}\bra{\boldsymbol{2}}-\ket{\boldsymbol{1}}\bra{\boldsymbol{2}}-\ket{\boldsymbol{2}}\bra{\boldsymbol{1}})\\
		&+\frac{1}{4}e^{-i\omega_0t}(-\ket{\boldsymbol{1}}\bra{\boldsymbol{0}}+\ket{\boldsymbol{2}}\bra{\boldsymbol{0}})\\
		&+\frac{1}{4}e^{i\omega_0t}(-\ket{\boldsymbol{0}}\bra{\boldsymbol{1}}+\ket{\boldsymbol{0}}\bra{\boldsymbol{2}}),
	\end{split}	
\end{equation}
and the concurrence entanglement of this state is equal to $\frac{1}{2}$, which is also invariant with time.

For a more general case where the coupling between two distant qubits is close to 1 but not exactly 1, we can also use the theory in the main text to give the quasi-steady state. For the two-qubit case investigated in this Appendix, the decoherence matrix can be written as
\begin{equation}
	\boldsymbol \Gamma=
	\begin{pmatrix} 
		1 & \alpha \\
		\alpha & 1 \end{pmatrix}
\end{equation}
with $\alpha$ being the incoherent coupling between two qubits. The eigenvalues of this matrix are $\Gamma_{\pm}=1\pm\alpha$, and the corresponding normalized eigenvectors are $\frac{1}{\sqrt{2}}(1,\pm 1)^\top$. Therefore, the collective $\nu$-operators are 
\begin{equation}
	\hat{\mathcal{O}}_{1}=\frac{1}{\sqrt{2}}\left( \hat{\sigma}_{1}+\hat{\sigma}_{2}\right), \hat{\mathcal{O}}_{2}=\frac{1}{\sqrt{2}}\left( \hat{\sigma}_{1}-\hat{\sigma}_{2}\right),
\end{equation}
and the master equation arrives at
\begin{equation}
    \begin{split}
        \frac{\partial \rho}{\partial t}&=\frac{i}{\hbar}[\rho, H]+\frac{\Gamma_{+}}{2} \left(2 \hat{\mathcal{O}}_{1} \rho \hat{\mathcal{O}}_{1}^{\dagger}-\rho \hat{\mathcal{O}}_{1}^{\dagger} \hat{\mathcal{O}}_{1}-\hat{\mathcal{O}}_{1}^{\dagger} \hat{\mathcal{O}}_{1} \rho\right)\\
        &+\frac{\Gamma_{-}}{2} \left(2 \hat{\mathcal{O}}_{2} \rho \hat{\mathcal{O}}_{2}^{\dagger}-\rho \hat{\mathcal{O}}_{2}^{\dagger} \hat{\mathcal{O}}_{2}-\hat{\mathcal{O}}_{2}^{\dagger} \hat{\mathcal{O}}_{2} \rho\right)\\
        &=\mathcal{L}\rho.
    \end{split}
\end{equation}
Following the same procedure above, we can find the eigenvalues and eigenvectors of the $\mathcal{L}$ matrix. If the value of $\alpha$ is close to 1, there exist four slow decay eigenvalues of $\mathcal{L}$ matrix, $L_0=0$, $L_{\alpha}=\alpha-1$, and $L_{\pm}=\frac{1}{2}(\alpha-1\pm 2i(\omega_0-g))$. The corresponding normalized and orthogonalized eigenvectors for these eigenvalues are the same with Eq.~(A5). Therefore, for a long period and the coherent coupling strength $g$, we can quickly obtain the state of the system
\begin{equation}
	\begin{split}
		\ket{r(t)}&= c_0\ket{\overline{l}_2}\ +\bra{\overline{l}_1}r(0) \rangle e^{L_{\alpha}t} \ket{\overline{l}_1}\\
		&+\bra{\overline{l}_3}r(0)\rangle e^{L_{-}t}\ket{\overline{l}_3}\\
		&+\bra{\overline{l}_4}r(0)\rangle e^{L_{+}t}\ket{\overline{l}_4} ,
	\end{split}
\end{equation}
with $c_0=1-\bra{\overline{l}_1}r(0) \rangle e^{L_{\alpha}t}$.

Here we will take two typical states as the initial states to clarify the dependence of the coherent coupling on our theory and the exact dynamics. The first state is the maximally entangled state, $\frac{1}{\sqrt{2}}(\ket{eg}-\ket{ge})$. We have $\bra{\overline{l}_1}r(0)\rangle=1$ and $\bra{\overline{l}_3}r(0)\rangle=\bra{\overline{l}_4}r(0)\rangle =0$, thus the final state according to Eq.~(A13) is given by $e^{L_{\alpha}t} \ket{\overline{l}_1}+(1-e^{L_{\alpha}t})\ket{\overline{l}_2}\
$, and its matrix form is 
\begin{equation}
	\rho_{\rm f}=\frac{1}{2}
	\begin{pmatrix}
		2-2e^{L_{\alpha}t} & 0 & 0 & 0  \\
		0 & e^{L_{\alpha}t} & -e^{L_{\alpha}t} & 0  \\
		0 & -e^{L_{\alpha}t} & e^{L_{\alpha}t} & 0  \\
		0 & 0 & 0 & 0  \\
	\end{pmatrix}.
\end{equation}
It can be verified that this state is the exact solution of the Lindblad master equation.
Two-qubit entanglement can be measured with entanglement concurrence \cite{wk98prl},  $C(\rho)={\rm max}\{0,\sqrt{\lambda_1}-\sqrt{\lambda_2}-\sqrt{\lambda_3}-\sqrt{\lambda_4}\}$ with $\lambda_i$s being the square roots of the eigenvalues of matrix $\rho \tilde{\rho}$, in decreasing order, and $\tilde{\rho}=(\sigma_y\otimes \sigma_y)\rho^*(\sigma_y\otimes \sigma_y)$. Therefore, the two-qubit concurrence of the theoretical final state shown in Eq.~(A14) can be calculated as $C_{12}=e^{(\alpha-1)t}$, which is slowly decaying over time when $\alpha$ is close to 1. Consequently, we know that under the maximally entangled and steady state of the two-qubit system, coherent has no influence on the final state.

\begin{figure}
	\includegraphics[width=0.5\textwidth]{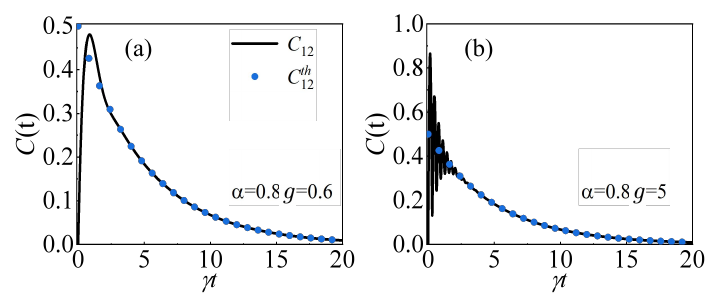}
	\caption{(Color online) The evolution of concurrence over time for the $\ket{eg}$ state under different coherent couplings, analytical results (black solid line), and results given by our theory (blue scattered points). (a) and (b) correspond to coherent couplings \textit{g}=0.6 and \textit{g}=5, respectively.}
\end{figure}

For another typical state $\ket{eg}$, the analytical form of its density matrix over time can be expressed as \cite{ren17jcp}
\begin{equation}
	\rho(t)=\frac{1}{4}
	\begin{pmatrix}
         4-2E_1^t & 0 & 0 & 0 \\
         0 & E_1^t-2E_3^t &  E_2^t-2iE_4^t & 0 \\
         0 & E_2^t+2iE_4^t & E_1^t+2E_3^t & 0 \\
         0 & 0 & 0 & 0 \\
	\end{pmatrix}
\end{equation}
where $E_1^t=e^{(\alpha-1)t}+e^{(-\alpha-1)t}$, $E_2^t=e^{(-\alpha-1)t}-e^{(\alpha-1)t}$, $E_3^t=2{\rm cos}(2gt)e^{-t}$, and $E_4^t=2{\rm sin}(2gt)e^{-t}$,
and the exact two-qubit concurrence of the state can be calculated as
\begin{equation}
    C_{12}^{}=\frac{1}{2}\sqrt{e^{2(\alpha-1)t}+e^{-2(\alpha+1)t}-2e^{-2t}{\rm cos}(4gt)}.
\end{equation}
According to Eq.~(A13), we have $\bra{\overline{l}_1}r(0)\rangle=\frac{1}{2}$ and $\bra{\overline{l}_3}r(0)\rangle=\bra{\overline{l}_4}r(0)\rangle =0$, thus the theoretical final state is $\frac{1}{2}e^{L_{\alpha}t} \ket{\overline{l}_1}+(1-\frac{1}{2}e^{L_{\alpha}t})\ket{\overline{l}_2}\
$, its matrix from is 
\begin{equation}
	\rho_{\rm f}=\frac{1}{4}
	\begin{pmatrix}
		4-2e^{L_{\alpha}t} & 0 & 0 & 0  \\
		0 & e^{L_{\alpha}t} & -e^{L_{\alpha}t} & 0  \\
		0 & -e^{L_{\alpha}t} & e^{L_{\alpha}t} & 0  \\
		0 & 0 & 0 & 0  \\
	\end{pmatrix},
\end{equation}
and it's concurrence can be calculated as 
\begin{equation}
    C_{12}^{th}=\frac{1}{2}e^{(\alpha-1)t}.
\end{equation}
Comparing these two results, we find that the exact concurrence shown in Eq.~(A16) will turn to the theoretical one shown in Eq.~(A18) at a large time $t$. However, the large coherent coupling $g$ can lead to the wild oscillation at the beginning of the dynamic. This can be clearly observed in Fig.~4.
We can also compare the similarity of density matrices between the exact and theoretical results by calculating the fidelity \cite{nielsen00book} 
\begin{equation}
    F(\rho(t),\rho_{\rm f})=\text{Tr}  \left[{\sqrt{\sqrt{\rho(t)} \rho_{\rm f} \sqrt{\rho(t)}}}\right],
\end{equation}
Substituting Eqs.~(A15) and (A17) into Eq.~(A19) we can obtain
\begin{equation}
\begin{split}
    F(\rho(t),\rho_{\rm f})&=\frac{1}{2}(e^{(\alpha-1)t}\\
    &+\sqrt{e^{2(\alpha-1)t}+e^{-2t}-4e^{(\alpha-1)t}-2e^{-(\alpha+1)t}+4}).
\end{split}
\end{equation}
According to Eq.~(A20), we can find that the fidelity is only related to incoherent coupling, and as time increases, the fidelity rapidly approaches 1.

\section{Three-qubit case and the orthogonalization procedure of steady basis}
In this Appendix, we first solve the steady basis of the $\mathcal{L}$ matrix of the three-qubit system and give theoretical results for its approximate steady state in the quasi-all-to-all case within the single-excitation basis. Then, we will present a general orthogonalization procedure on the steady basis and an example in this three-qubit case. We find that five eigenvalues of the three-qubit case are equal to zero on a single-excitation basis. Except for the trivial vacuum state $\ket{l_0}$, the remaining four eigenvectors with zero eigenvalues are marked as $\ket{l_1}\sim\ket{l_4}$. Here we define $\ket{l_i}=\tilde{a}_i$, and their corresponding matrix form $a_i$ for convenience can be written as
\begin{equation}
	\begin{split}
	&a_1=\frac{1}{2}(\ket{\boldsymbol{1}}-\ket{\boldsymbol{2}})\otimes(\bra{\boldsymbol{1}}-\bra{\boldsymbol{2}}),\\
	&a_2=\frac{1}{2}(\ket{\boldsymbol{1}}-\ket{\boldsymbol{3}})\otimes(\bra{\boldsymbol{1}}-\bra{\boldsymbol{3}}),\\
    &a_3=\frac{1}{2}(\ket{\boldsymbol{1}}\bra{\boldsymbol{1}}+\ket{\boldsymbol{2}}\bra{\boldsymbol{3}}-\ket{\boldsymbol{2}}\bra{\boldsymbol{1}}-\ket{\boldsymbol{1}}\bra{\boldsymbol{3}}),\\
    &a_4=\frac{1}{2}(\ket{\boldsymbol{1}}\bra{\boldsymbol{1}}+\ket{\boldsymbol{3}}\bra{\boldsymbol{2}}-\ket{\boldsymbol{3}}\bra{\boldsymbol{1}}-\ket{\boldsymbol{1}}\bra{\boldsymbol{2}}).
 \end{split}
\end{equation}

Within the single-excitation basis, there are four pure oscillation terms, $\epsilon_5=\epsilon_6=-i\omega_0$ and $\epsilon_7=\epsilon_8=i\omega_0$, and their corresponding eigenvectors are marked as $\ket{l_5}\sim\ket{l_8}$ with the corresponding matrices being
\begin{equation}
\begin{split}
	&b_1=\frac{1}{\sqrt{2}}(\ket{\boldsymbol{2}}\bra{\boldsymbol{0}}-\ket{\boldsymbol{1}}\bra{\boldsymbol{0}}),\\
	&b_2=\frac{1}{\sqrt{2}}(\ket{\boldsymbol{3}}\bra{\boldsymbol{0}}-\ket{\boldsymbol{1}}\bra{\boldsymbol{0}}),\\
	&b_3=\frac{1}{\sqrt{2}}(\ket{\boldsymbol{0}}\bra{\boldsymbol{2}}-\ket{\boldsymbol{0}}\bra{\boldsymbol{1}}),\\
	&b_4=\frac{1}{\sqrt{2}}(\ket{\boldsymbol{0}}\bra{\boldsymbol{3}}-\ket{\boldsymbol{0}}\bra{\boldsymbol{1}}).\\
    \end{split}
\end{equation}

Observations indicate that these states are not orthogonal to each other. Here, we take the Gram-Schmidt process \cite{david97book} to orthogonalize the zero-eigenvalue eigenvectors. The Gram-Schmidt process is as follows: given $k$ vectors $v_1,\cdots,v_k$, their orthogonalized vectors can be calculated as
\begin{equation}
	\begin{split}
	u_1 &= v_1, \\
	u_2 &= v_2 - \mathrm{proj}_{u_1}(v_2), \\
	u_3 &= v_3 - \mathrm{proj}_{u_1}(v_3) - \mathrm{proj}_{u_2}(v_3), \\
	u_4 &= v_4 - \mathrm{proj}_{u_1}(v_4) - \mathrm{proj}_{u_2}(v_4) - \mathrm{proj}_{u_3}(v_4), \\
	&\vdots \\
	u_k &= v_k - \sum_{j=1}^{k-1} \mathrm{proj}_{u_j}(v_k).
    \end{split}
\end{equation}
Among the above vectors, $\mathrm{proj}_{a}(b)$ represents the orthogonal projection of a vector $b$ onto a vector $a$.
\begin{equation}
	\mathrm{proj}_{a}(b)=\frac{\langle b,a\rangle}{\langle a, a \rangle} a,
\end{equation}
and $\langle b,a\rangle$ is the inner product of vectors $a$ and $b$.
After flattening $a_2$, $a_3$, $a_4$, and $a_1$ as the vectors $v_1$, $v_2$, $v_3$, and $v_4$, respectively, according to the procedure shown in Eq.~(B3), we can obtain four orthogonal vectors $u_1$, $u_2$, $u_3$, and $u_4$. Normalizing these vectors we can obtain the orthogonalized and normalized vectors, and their corresponding matrix form $A_1\sim A_4$ can be written as
\begin{equation}
\begin{split}
	A_1=&\frac{1}{2}(\ket{\boldsymbol{1}}-\ket{\boldsymbol{3}})\otimes(\bra{\boldsymbol{1}}-\bra{\boldsymbol{3}}),\\
	A_2=&\frac{1}{2\sqrt{3}}[(\ket{\boldsymbol{1}}+\ket{\boldsymbol{3}})\otimes(\bra{\boldsymbol{1}}-\bra{\boldsymbol{3}})+2(\ket{\boldsymbol{2}}\bra{\boldsymbol{3}}-\ket{\boldsymbol{2}}\bra{\boldsymbol{1}})],\\
	A_3=&\frac{1}{2\sqrt{3}}[(\ket{\boldsymbol{1}}-\ket{\boldsymbol{3}})\otimes(\bra{\boldsymbol{1}}+\bra{\boldsymbol{3}})+2(\ket{\boldsymbol{3}}\bra{\boldsymbol{2}}-\ket{\boldsymbol{1}}\bra{\boldsymbol{2}})],\\
	A_4=&\frac{1}{6}[(\ket{\boldsymbol{1}}+\ket{\boldsymbol{3}})\otimes(\bra{\boldsymbol{1}}+\bra{\boldsymbol{3}})+4\ket{\boldsymbol{2}}\bra{\boldsymbol{2}}-2(\ket{\boldsymbol{2}}\bra{\boldsymbol{1}}\\
&+\ket{\boldsymbol{2}}\bra{\boldsymbol{3}}+\ket{\boldsymbol{1}}\bra{\boldsymbol{2}}+\ket{\boldsymbol{3}}\bra{\boldsymbol{2}})].
	\end{split}	
\end{equation}
Similarly, following the same orthogonalization procedure, we can obtain the orthogonalized matrix form $B_1\sim B_4$ of Eq.~(B2)
\begin{equation}
\begin{split}
	&B_1=\frac{1}{\sqrt{2}}(\ket{\boldsymbol{2}}\bra{\boldsymbol{0}}-\ket{\boldsymbol{1}}\bra{\boldsymbol{0}}),\\
	&B_2=\frac{1}{\sqrt{6}}(-\ket{\boldsymbol{1}}\bra{\boldsymbol{0}}-\ket{\boldsymbol{2}}\bra{\boldsymbol{0}}+2\ket{\boldsymbol{3}}\bra{\boldsymbol{0}}),\\
	&B_3=\frac{1}{\sqrt{2}}(\ket{\boldsymbol{0}}\bra{\boldsymbol{2}}-\ket{\boldsymbol{0}}\bra{\boldsymbol{1}}),\\
	&B_4=\frac{1}{\sqrt{6}}(-\ket{\boldsymbol{0}}\bra{\boldsymbol{1}}-\ket{\boldsymbol{0}}\bra{\boldsymbol{2}}+2\ket{\boldsymbol{0}}\bra{\boldsymbol{3}}),
    \end{split}	
\end{equation}
where $\tilde{A}_1$ to $\tilde{A}_4$ are the eigenstates correspond to the eigenvalue 0,  $\tilde{B}_1$ and  $\tilde{B}_2$ correspond to the eigenvalue $-i\omega_0$, and  $\tilde{B}_3$ and  $\tilde{B}_4$ correspond to the eigenvalue $i\omega_0$.

For systems with more qubits, the orthogonalized steady-state basis vectors can also be easily obtained according to the above process as well as Eqs.~(25) and (26) in the main text. Based on these orthogonalized basis vectors, we can easily obtain the approximate final state of the system.

% Here we give an example. For the initial state $\rho(0)=\ket{gge}$, the nontrivial inner product of $\bra{\overline{l}_i}r(0)\rangle$ can be easily calculated as $\bra{\overline{l}_1}r(0)\rangle=\frac{1}{2}$, $\bra{\overline{l}_2}r(0)\rangle=\bra{\overline{l}_3}r(0)\rangle=\frac{1}{2\sqrt{3}}$, $\bra{\overline{l}_4}r(0)\rangle=\frac{1}{6}$ and $\bra{\overline{l}_5}r(0)\rangle=\bra{\overline{l}_6}r(0)\rangle=\bra{\overline{l}_7}r(0)\rangle=\bra{\overline{l}_8}r(0)\rangle=0$. Thus, for an initially roduct state $\ket{gge}$, the normalized final steady state will be
% \begin{equation}
% 	\ket{r_{\rm s}}=\frac{1}{3}\ket{\overline{l}_0}+\frac{1}{2} \ket{\overline{l}_1}+\frac{1}{2\sqrt{3}}\ket{\overline{l}_2}+\frac{1}{2\sqrt{3}}\ket{\overline{l}_3}+\frac{1}{6}\ket{\overline{l}_4},
% \end{equation}
% which is a three-qubit entangled state with the three-qubit negativity being $N_3=0.298$. Therefore, we can readily obtain the exact final steady state for a fixed initial state by avoiding complex dynamic calculations. 

\begin{figure}
	\includegraphics[width=0.5\textwidth]{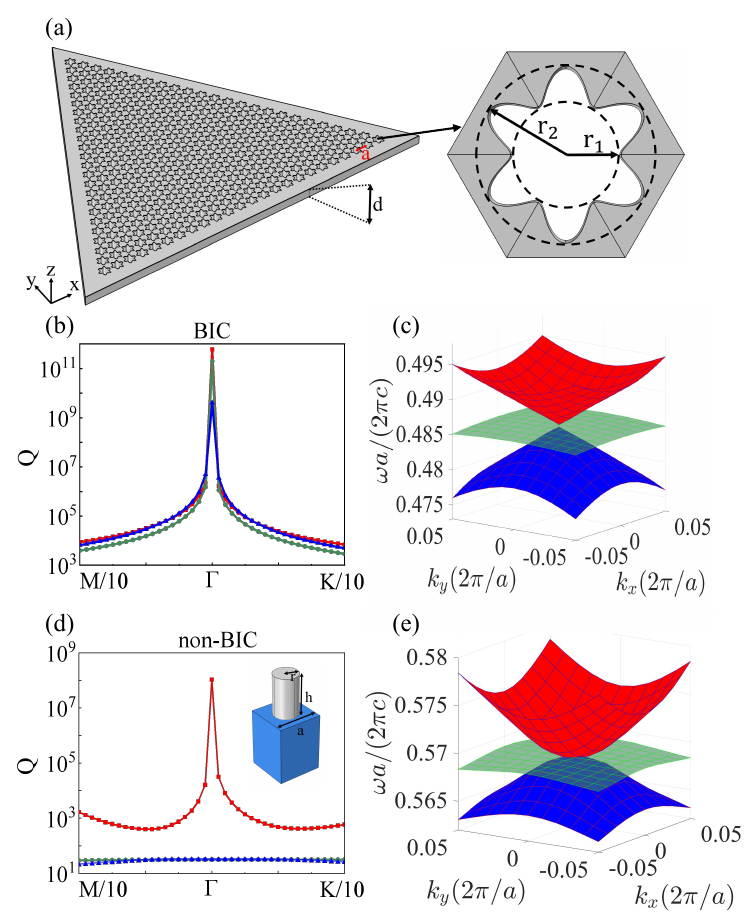}
	\caption{(Color online) (a) is a schematic diagram of the BIC photonic crystal slab model, with the right figure showing a detailed diagram of the structure within a single unit cell. (b) and (c) represent the quality factor of the Dirac cone and the band diagram, respectively. (d) and (e) are the quality factor and band diagram corresponding to the non-BIC photonic crystal slab, respectively. The inset in Figure (d) shows a schematic diagram of its unit cell structure.}
\end{figure}

%\section{The orthogonalization process of steady-state basis}
%Here, we take the Gram-Schmidt process \cite{david97book} to orthogonalize the zero-eigenvalue eigenvectors. The Gram-Schmidt process is as follows.

%Given $k$ vectors $v_1,\cdots,v_k$, their orthogonalized vectors can be calculated as
%\begin{equation}
%	\begin{split}
%	u_1 &= v_1, \\
%	u_2 &= v_2 - \mathrm{proj}_{u_1}(v_2), \\
%	u_3 &= v_3 - \mathrm{proj}_{u_1}(v_3) - \mathrm{proj}_{u_2}(v_3), \\
%	u_4 &= v_4 - \mathrm{proj}_{u_1}(v_4) - \mathrm{proj}_{u_2}(v_4) - \mathrm{proj}_{u_3}(v_4), \\
%	&\vdots \\
%	u_k &= v_k - \sum_{j=1}^{k-1} \mathrm{proj}_{u_j}(v_k),
%    \end{split}
%\end{equation}
%where $\mathrm{proj}_{a}(b)$ represents the orthogonal projection of a vector $b$ onto a vector $a$.
%\begin{equation}
%	\mathrm{proj}_{a}(b)=\frac{\langle b,a\rangle}{\langle a, a \rangle} a,
%\end{equation}
%and $\langle b,a\rangle$ is the inner %product of vectors $a$ and $b$.

\section{Calculations of the bound state in the continuum}
In the main text, we selected a photonic crystal slab with BIC characteristics to verify our theory, as shown in Fig.~5(a). The photonic crystal slab is made up of silicon slab of thickness $d=0.5a$ that has air holes arranged in a triangular lattice with a lattice constant $a$, and in actual calculations, we used a triangular sample slab.
The shape of the air holes [as shown in the right inset of Fig.~5(a)] is given as $r(\theta)=\frac{1}{2}(r_2+r_1)+\frac{1}{2}(r_2-r_1){\rm cos}(6\theta)$ in polar coordinates, where $r_1 = 0.27a$ and $r_2 = 0.43a$. Fig.~5(b) and 5(c) represent the quality factors and band structures of three degenerate bands near the $\Gamma$ point, respectively \cite{minkov18prl}.

In contrast, we have chosen another photonic crystal structure as shown in the inset of Fig.~5(d) that does not possess BIC characteristics. The non-BIC photonic crystal slab is composed of silicon pillars arrayed in a square lattice on a silicon dioxide substrate, with a height of $h=860~\rm nm$, a radius of $r=256~\rm nm$, and a lattice constant of $a=918~\rm nm$ \cite{kita17oe}. Its band structure still exhibits degeneracy in the center of the Brillouin zone, as shown in Fig.~5(e), but only the monopole mode has a high-quality factor due to symmetry, while two of the degenerate modes have a relatively low-quality factor; therefore, Dirac point modes do not exhibit BIC characteristics, as shown in Fig.~5(d). In this case, although the effective refractive index is still near zero when the light is transmitted in the plane of the slab, the leakage perpendicular to the slab direction will also be very large. 

\begin{figure}
	\includegraphics[width=0.5\textwidth]{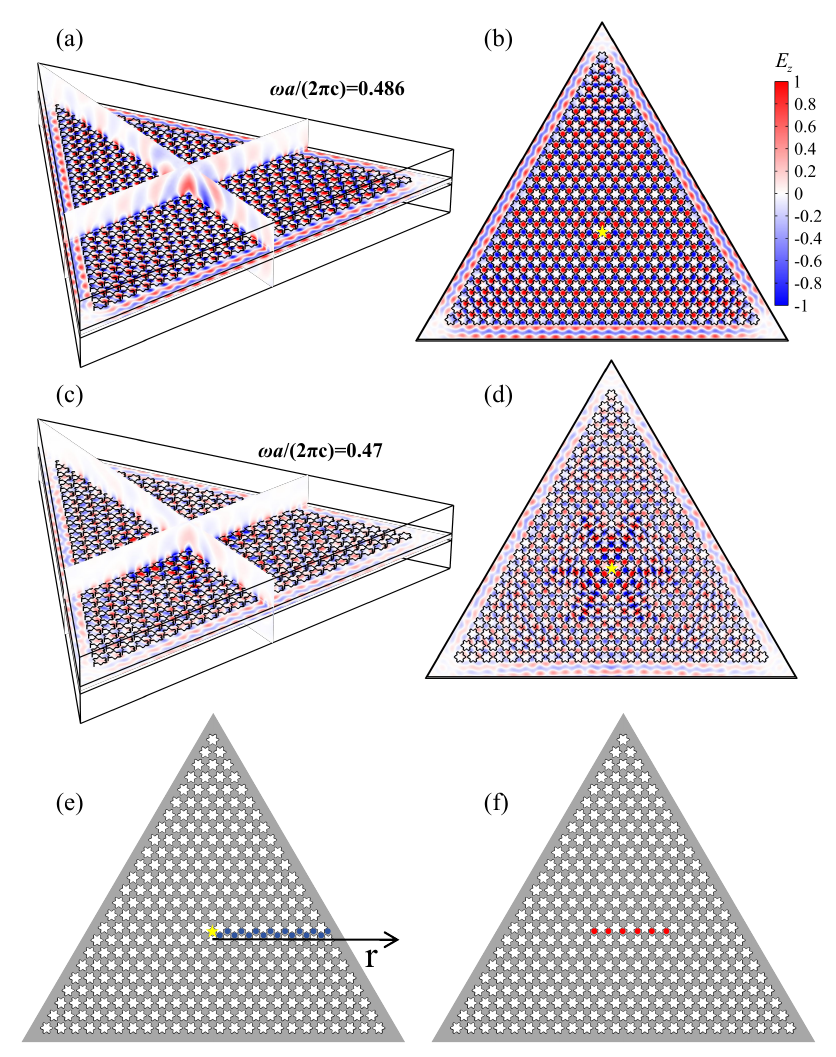}
	\caption{(Color online)  (a) and (b) represent the three-dimensional view and the cross-sectional electric field distribution within the slab, with the emitter (yellow star) located at the geometric center of the slab. (a) and (b) are the electric field diagrams corresponding to the Dirac point frequency, while (c) and (d) are the electric field diagrams for frequencies other than the Dirac point. (e) is a schematic diagram of the calculation positions for coherent and incoherent coupling, with yellow stars indicating the positions of the emitters and blue dots representing the calculation positions. (f) is a schematic diagram of the positions of the emitters (red dots) for the calculation of steady-state evolution in the BIC photonic crystal slab.}
\end{figure}

Fig.~6 presents our simulated calculations of the electric field excited by emitters (yellow star) at different frequencies embedded in the photonic crystal slab. The electric field intensity distribution diagram shows the normalized $E_z$ when $a=800~\rm nm$. Figure~6(a) and 6(b) show the three-dimensional view and the cross-sectional field distribution within the slab under the BIC frequency $\omega a/(2\pi c) = 0.486$, respectively; while Fig.~6(c) and 6(d) correspond to the non-Dirac case of $\omega a/(2\pi c )= 0.47$. Comparing the field distributions under BIC and non-BIC cases, it can be seen that the field distribution in the slab under BIC is very uniform at high symmetry positions, even at the edge of the slab sample, while the field distribution at high symmetry points decreases slightly under non-BIC frequencies. In addition, due to the coupling between the emitter and the high-quality BIC, the radiation leakage in the z direction is also limited.
%\begin{figure}
	%	\epsfig{figure=fig4.pdf,width=0.5\textwidth}
	%\includegraphics[width=0.5\textwidth]{fig7.pdf}
	%\caption{(Color online) (a) is a schematic diagram of the calculation positions for coherent and incoherent coupling, with yellow stars indicating the positions of the emitters and blue dots representing the calculation positions. (b) and (c) depict the measured coherent and incoherent coupling strengths at these locations for photonic crystal slabs of different sizes. (d) is a schematic diagram of the positions of the emitters (red dots) for the calculation of steady-state evolution in the BIC photonic crystal slab. (e) and (f) present the coherent and incoherent coupling matrices measured at these six positions.}
%\end{figure}
Figure~6(e) illustrates the schematic of the emitter arrangement for pairwise coherent and incoherent coupling on the BIC photonic crystal slab, as shown in Fig.~3(c) and 3(d) of the main text. In Fig.~6(e), the yellow star marks the position of the fixed emitter, and the remaining blue dots indicate the positions of the other emitter. The red dots in Fig.~6(f) present the emitter arrangement of six emitters referred in Fig.~3(e) of the main text. The evolution results are shown in Fig.~3(e). The simulations of the couplings and the field patterns in this work are implemented through the software COMSOL multiphysics.

%%%%%%%%%%%%%%%%%%%%%%%%%%%%%%%%%%%%%%%%%%%%%%%%%%%%%%%%%%%%%%%%%%%%%%%%%%%%%%%%%%%%

% \bibliography{apssamp}% Produces the bibliography via BibTeX.

\end{document}